# An Advanced Survey on Secure Energy-Efficient Hierarchical Routing Protocols in Wireless Sensor Networks

Abdoulaye Diop[1], Yue Qi[2], Qin Wang[3] and Shariq Hussain[4]

[1,2,3,4] School of Computer and Communication Engineering, University of Science and Technology Beijing
Beijing, 100083, China
*b20100556@xs.ustb.edu.cn, qiyuee@ustb.edu.cn, wangqin@ies.ustb.edu.cn, b20100557@xs.ustb.edu.cn*

**Abstract**

Wireless Sensor Networks (WSNs) are often deployed in hostile environments, which make such networks highly vulnerable and increase the risk of attacks against this type of network. WSN comprise of large number of sensor nodes with different hardware abilities and functions. Due to the limited memory resources and energy constraints, complex security algorithms cannot be used in sensor networks. Therefore, it is necessary to balance between the security level and the associated energy consumption overhead to mitigate the security risks. Hierarchical routing protocol is more energy-efficient than other routing protocols in WSNs. Many secure cluster-based routing protocols have been proposed in the literature to overcome these constraints. In this paper, we discuss Secure Energy-Efficient Hierarchical Routing Protocols in WSNs and compare them in terms of security, performance and efficiency. Security issues for WSNs and their solutions are also discussed.

## 1. Introduction

The tremendous development in the electronics technology lead the way to development of micro-electronics thus enabling production of small chips and micro devices. The communication technology is being reformed due the design and development of micro devices and hence enabled the design and development of WSNs with low cost, low energy consumption and high utilization. WSNs have lot of applications in military, health and other industrial sectors. Because of the characteristics of WSNs, sensor nodes are usually characterized by limited power, low bandwidth, memory size and limited energy [1].

Due to the scalability and energy efficiency characteristics, researchers proposed many routing protocols for cluster-based WSNs [2]. In WSNs, routing protocols can be classified into two categories: Network Structure and Protocol Operation. Hierarchal routing protocol is one of the categories in their classification of WSN routing protocol based on the Network Structure. In cluster-based routing protocols, network is divided into cluster and each cluster has its own cluster head (CH). Further, CHs are responsible for relaying of messages from ordinary nodes to the Base Station (BS). CHs can communicate directly with the BS, can be anywhere in the network and change per interval, which also improves network's energy efficiency [2].

Several enhanced secure hierarchal routing protocols have been proposed in literature [19, 21, 23, 24-28], to attempt to achieve both security and efficiency for WSNs. Most routing protocols are vulnerable to a number of security threats [3]. Attacks involving CHs are the most damaging.

Due to the resource constraints of wireless sensors, public-key based cryptographic algorithms like RSA and Diffie-Hellman are too complicated and energy-consuming for WSNs. However the symmetric cryptographic technique has its own qualities that always make it more favourite as compared to public key cryptography for WSNs. Furthermore to provide security in WSN, encryption keys must be established among sensor nodes. Key distribution refers to the distribution of multiple keys among the sensor nodes. Key management also receives a great deal of attention in data encryption and authentication in WSNs security.

Hence, it is necessary to well balance security level and the associated energy consumption overhead, to mitigate the security risks. Keys that are necessary for security and efficiency requirements of WSNs are listed in Table1.

In this paper, we present an Advanced Survey on Secure Energy-Efficient Hierarchical Routing Protocol in WSNs. Security issues are discussed and their solutions presented.

The rest of the paper is organized as follows. Section II describes the Energy constraints in WSNs while Section III presents a review of energy-efficient hierarchical cluster routing protocols. Security requirements in WSNs are presented in Section IV. Section V discusses various attacks that can be launched on routing protocols in WSNs.

Table 1: Design requirement of energy-efficient security scheme

| S. No | *Requirement Type* | Requirements |
|---|---|---|
| 1. | Security Requirement | Authentication<br>Secrecy<br>Integrity<br>Resilience against node capture<br>Resistance against node replication<br>Compromised node revocation<br>Fresh node addition |
| 2. | Efficiency Requirement | Energy efficiency<br>Network connectivity<br>Maximum supported network size<br>Minimum memory storage<br>Low computational overhead<br>Low communication overhead |

Basic security mechanisms in WSNs are presented in section VI. In Section VII, secure hierarchical routing protocols in WSNs are discussed. Security analysis is presented in Section VIII and finally Section IX concludes our work.

## 2. Energy constraints in WSNs

The biggest constraint among the rest of the major constraints of a WSN is energy. In most cases the battery replacement is impossible. This means that the lifetime of a sensor depends greatly on the life of the battery. Fig. 1 illustrates sensor node architecture with four major components and associated energy cost parameters [2]. Basically, energy consumption in sensor nodes can be classified into following three parts, as shown in Fig. 2.

In WSNs, communication is more costly than computation [4] and many operations are energy intensive. It is for this reason that the current research focuses primarily on ways to reduce energy consumption.

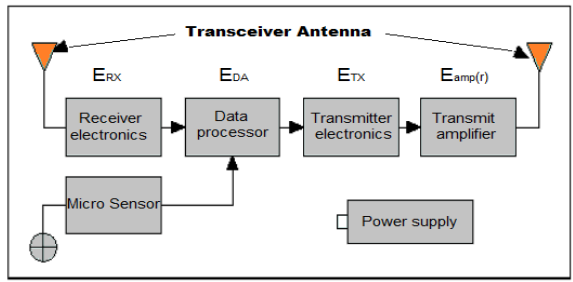

Fig. 1 Major components and associated energy cost parameters of a sensor node.

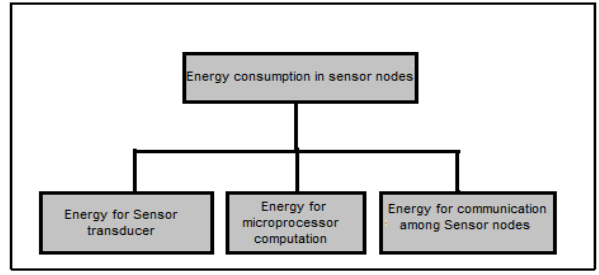

Fig. 2 Sensor nodes energy consumption.

In WSNs, the security mechanisms (e.g., encryption, decryption, signing data, verifying signatures) are the main factors that influence power consumption by sensor nodes. Hence, adding security to WSNs also impose overhead to power consumption, the energy required to store security parameters and the energy required to transmit the security parameters. Limited energy prohibits the use of complex security mechanisms for message expansion. Furthermore, in WSNs security mechanism usually use more energy consumption for higher security levels. Thus, WSNs could be classified into different security levels following energy cost [5].

## 3. Energy-Efficient Hierarchical routing Algorithms in WSNs

HWSNs is one of the main research areas in WSNs and behave the most energy-efficient among the rest of protocols for WSNs. Table 2 shows routing protocols classification in WSNs regarding to different categories.

Many research projects during the last few years have explored cluster based routing protocols in WSNs from different perspectives. Most of them have been proposed for routing the correct data to the BS and prolonging the life of sensors node. Hence, each protocol has advantages and disadvantages.

Table 2: Routing protocols classification in WSNs

| *Routing Protocols in WSNs* | Network Structure | Flat network routing<br>Hierarchical<br>Location Based-routing |
|---|---|---|
| | Protocol Operation | Negotiation-based routing<br>Multipath-based routing<br>Query-based routing<br>QoS-based routing<br>Coherent-based routing |

## 3.1 Low Energy Adaptive Clustering Hierarchy (LEACH).

LEACH [2], first energy-efficient hierarchical routing protocol, is proposed for WSN using homogenous stationary nodes. In LEACH, Sensors nodes choose their leader based on some parameters such as the strongest signal received from a CH. After certain interval, new nodes are selected as CH. LEACH reduces energy consumption by utilizing randomize rotation of CHs to evenly distribute the energy load in the network and turning off ordinary nodes when not required.

## 3.2 Power-Efficient Gathering in Sensor Information Systems (PEGASIS)

PEGASIS [6] is an extension of LEACH protocol. PEGASIS forms chains from sensors nodes. Sensors nodes transmit or receive data from a neighbor, in this way PEGASIS avoids cluster formation and uses only one node in a chain to transmit to the base-station. Therefore, increases the network lifetime.

## 3.3 Hybrid Energy-Efficient Distributed Clustering (HEED).

HEED [7] is an improvement of LEACH. HEED clustering randomly selects CHs and improves the lifetime of the network over LEACH clustering.

## 3.4 Energy-Efficient Homogeneous Clustering Algorithm (EEHCA) for Wireless Sensor Networks

In EEHCA [8], a new CH is selected based on the residual energy of existing CH, nearest hop distance of the node and holdback value. The uniform distribution of the cluster members extended the network lifetime.

## 4. Security Requirements in WSNs

To address the security issues in WSNs, we come across certain security requirements that must be addressed in WSNs environment. Here are some core security properties, implementation of which can contribute in development of more secure WSNs.

### 4.1 Authentication

It enables entities to cooperate within WSN without risk, by identifying and controlling participants in the network. It appears to be the cornerstone of a WSN. We cannot ensure confidentiality and the integrity of exchanged messages, if from the start we are not sure to communicate with the correct nodes. Therefore, it is essential for a receiver to have a mechanism to verify that the received packets have indeed come from the actual sender node. We can use Message Authentication Code (MAC) to ensure both the authentication of the origin of the message and integrity. An example of MAC is HMAC [9].

### 4.2 Data Integrity

It ensures that no message can be altered by an entity as it traverses from the sender to the recipient. It can be ensured by the use of cryptographic hash functions, which require obtaining a fingerprint for each digital message. MD5 function and Secure Hash Algorithm-1 (SHA-1) [10] are some examples of most used hash functions.

### 4.3 Data Confidentiality

Once the message parts are authenticated, confidentiality remains an important point. It is to keep the secrecy of exchanged messages. The confidentiality can be ensured by the use of cryptography keys (i.e. symmetric or asymmetric).

### 4.4 Availability

It ensures that the services of a WSN should be always available even in the presence of an internal or external attack. A central access control system is used to ensure successful delivery of every message to its recipient.

### 4.5 Data Freshness

This service ensures that the data is up-to-date and ensures that no adversary can replay old messages. Data freshness is important when the WSN nodes use shared keys for message communication. The risk is that a potential adversary can launch a replay attack using the old key, as the new key is being refreshed and propagated to all the nodes in the WSN. A nonce or time-specific counter may be added to each packet to check the freshness of the packet.

### 4.6 Self-organization

In a WSN, each node should be self-organizing. This requirement of WSN also poses a great challenge to security. The dynamic nature of a WSN makes it sometimes impossible to deploy any preinstalled shared key mechanism among the sensors nodes and the BS [11]. It is desirable that in WSNs, the nodes self-organize among themselves not only for multi-hop routing but also to carryout key management and developing trust relations.

## 5. Routing Attacks in WSNs

The network layer of WSNs suffers from different types of attacks such as: (i) Sybil, (ii) sinkhole, (iii) hello flood, (iv) wormhole, (v) selective packet forwarding, etc. These attacks are described briefly. Table 3 illustrates the routing attacks on WSNs and some solutions to defeat them.

### 5.1 Sybil Attack

The attacker presents multiple identities on one node in the network. In this way, the attacker mostly affects the routing mechanism. Generally Sybil attacks are prevented by validation techniques.

### 5.2 Sinkhole Attack

In this type of attack, attacker presents himself in the network with high capability resources, by which announces a short path to destination to attract packets and then may drop them [12]. In this way, sinkhole attack gives birth to some attacks like blackhole, selective forwarding, etc.

### 5.3 Hello Flood Attack

Strong hello message broadcasted by attacker with high transmission power is to be received by every node in the network [12]. Other nodes may think this message is nearest to them and sends packets by this node. In this way, attack congestion occurs in the network. Hello flood attacks are prevented using blocking techniques.

### 5.4 Wormhole Attack

An adversary launch wormhole with tunneling mechanism to establish him between entities by confusing the routing protocol. Using out-of-bound channel to route packets, makes this kind of attack very difficult to detect.

### 5.5 Selective Forwarding

Generally two factors are important in this attack. The first is location of attacker as it will attract more traffic if the location of malicious node is close to base. The number of dropped messages is another factor, the more messages drops, the more energy it has in order to attack. An adversary can selectively forward some messages and drops others, therefore may compromise a node [12].

## 6. Basic Security Mechanisms in WSNs

Security in sensor networks poses different challenges than conventional network, due to inherent resources and computing constraints. However, secure communications in some WSNs are critical. Two security aspects such as the area of cryptography and key management received a great deal of attention in WSNs. Cryptography and key management mechanisms for WSN security are presented below.

### 6.1 Key Management

Key management is the process in which keys are created, stored, protected, transferred, used between authorized parties and destroyed when they do not need [13]. Key management establishes the keys that are necessary to provide confidentiality, integrity and authentication services. Due to the limited memory resources and energy constraints of sensor nodes, complex security algorithms cannot be used in sensor networks. The main goal of key management in WSNs is to ensure security requirements of WSN by encrypting messages and authenticates the communicating nodes. Key management is quite challenging issue in WSNs and researchers presented a large number of approaches in literature due to the importance of key management in WSNs. Some researchers have investigated the WSNs key management schemes and divided them into different categories.

From the work of Xiangqian and Makki [14], key management schemes in WSNs can be classified as following: key pre-distribution schemes, hybrid cryptography schemes, one-way hash schemes and key management in hierarchy networks.

– Key Pre-Distribution Schemes: refers to how many keys are needed and how should the keys be distributed before the nodes are deployed? Key pre-distribution schemes can be classified as probabilistic schemes and deterministic schemes.

Table 3: Routing attacks on WSNs and countermeasures

| Layer | Attacks | Solutions |
|---|---|---|
| Network | Spoofed routing information & selective forwarding | Egress filtering, authentication, monitoring |
| | Hello Flood | Authentication, packet leashes by using geographic and temporal info |
| | Wormhole | Authentication, probing |
| | Sinkhole | Redundancy checking |
| | Sybil | Authentication, monitoring, redundancy |

In probabilistic scheme, the existence of a shared key between a particular pair of nodes is not certain and instead guaranteed only probabilistically. The basic idea of these schemes is to randomly preload each sensor with a subset of keys from a global key pool before deployment. So, these schemes can also be called Random Key Pre-distribution (RKP). The first probabilistic key pre-deployment scheme is introduced by Eschenauer and Gligor [11], which consists of three phases: key pre-distribution, shared-key discovery, and path-key establishment.

Contrary to probabilistic schemes, deterministic schemes guarantee that any two intermediate nodes can share one or more pre-distribution keys.

LEAP [15] (Localized Encryption and Authentication Protocol), is a basic example of deterministic key management scheme. The authors of LEAP establish four types of keys that must be stored in each sensor to ease the overhead of key management and to provide secure communications in WSNs.
– Hybrid Cryptography Schemes: use both asymmetric-key and symmetric-key cryptographs.
– One-way Hash Schemes: is used in many approaches that come from one-way hash function technique to ease key management.
– Key Management in Hierarchy Networks: many key management approaches are based on a normal flat structure. There are still some approaches that utilize a hierarchical structure in order to ease the difficulties by balancing the traffic among a BS, CHs, and sensors. These are the three parts of networks that have different resources.

Zhang and Varadharajan [1] also considered three important factors for classification of key management schemes in WSNs based on the encryption techniques. These include symmetric, asymmetric and hybrid. Based on the key establishment mechanism, Zhang and Varadharajan [1] divided the symmetric and asymmetric schemes into eight and three subcategories respectively. This classification is shown in Fig. 3.

– Most of the WSNs use the symmetric key schemes because these schemes requires less computation time than other schemes. Based on the key distribution, key discovery and key establishment in the schemes, symmetric schemes can be divided into six categories: entity based schemes, pure probabilistic-based schemes, polynomial-based key pre-distribution schemes, matrix-based key pre-distribution schemes, tree-based key

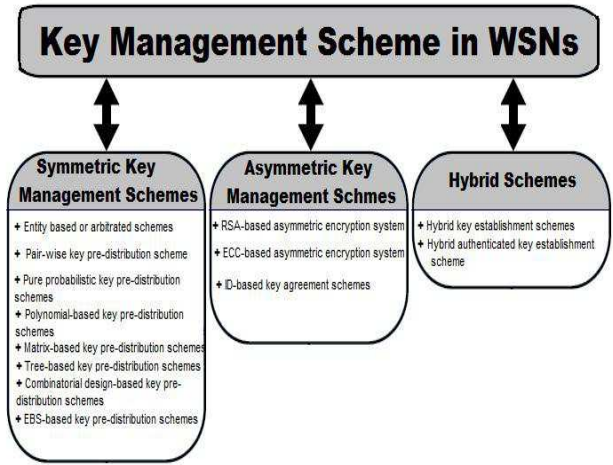

Fig. 3 Key management scheme in WSNs

pre-distribution schemes and EBS-based key pre-distribution schemes.

– In asymmetric key management schemes, RSA and Elliptic Curve Cryptography (ECC) are two major public key techniques. Public key technology is widely used in the security of Internet. On the other hand, some researchers believe that these techniques are too heavy-weight for sensor networks because of requirement constraints.

– In Hybrid schemes, several research groups [16] proposed the hybrid key establishment schemes for WSNs. The motivation is to exploit the difference among the BS, the CH and the sensor, and place the cryptographic burden on the BS or the sensors whose sources are less constrained. Sensors have limited computational power and energy resources, whereas BS has much more computational power and other resources. The hybrid key establishment schemes reduce the high computational cost on the sensors by placing them on the BS side.

## 6.2 Cryptographic Mechanisms

There are two types of cryptography techniques depending on the key. First is symmetric key cryptography that uses the same key for encryption and decryption (e.g., AES). Second is asymmetric key cryptography that uses different keys to encrypt and decrypt (e.g., RSA), requires more computation resources than symmetric key cryptography. Symmetric and Asymmetric encryption are illustrated in Fig. 4 (a) and (b) respectively. Due to the limited resources, public key cryptographic algorithms are not suitable for WSNs. However, the symmetric cryptographic technique has its own qualities that make it more favourable as compared to public key cryptography for WSNs.

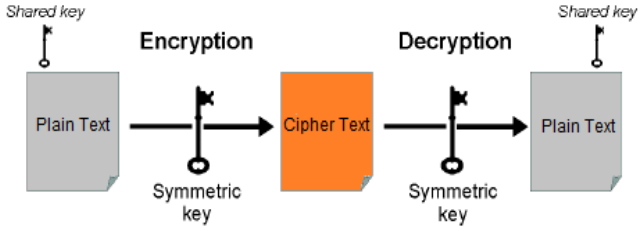

(a) Symmetric Encryption

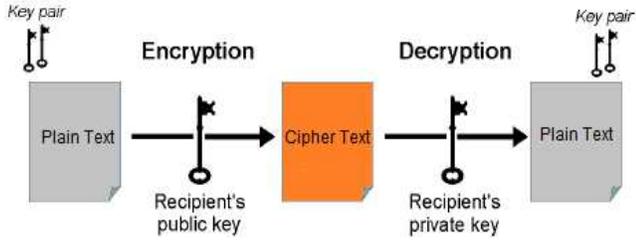

(b) Asymmetric Encryption

Fig. 4 Types of cryptography

For secure communication in WSNs, it is necessary to choose the most efficient cryptographic algorithm. A benchmark on cryptographic algorithms was presented by Law et al. [17] for WSNs. They focus on storage, energy-efficiency and security properties of Skipjack, RC5, RC6, and Rijndael.

In [17], the result of analysis show that in an environment where security is important, memory efficient cryptographic algorithm is required. And in an environment where availability of network is important, energy-efficient cryptographic algorithm has to be used. TinySec [18] provides Skipjack and RC5 as recommended cryptographic algorithm. Each algorithm has its own property, memory, energy efficiency and security. Table 4 summarized the execution time of ECC, RSA, RC5 and Skipjack implemented on an Atmega128 processor.

Table 4: Execution time of some cryptographic algorithms

| S. No | Algorithms | Operation Time |
|---|---|---|
| 1. | RC5 | 0.38 ms |
| 2. | Skipjack | 0.26 ms |
| 3. | ECC-160 | 810 ms |
| 4. | RSA-1024 | 10990 ms |

## 7. Secure Hierarchical Routing Protocols in WSNs

Security in cluster based routing protocols is a particularly challenging task. Many works have been proposed to secure the hierarchical routing protocols. In this section, some approaches are reviewed and compared. Their advantages and disadvantages are also discussed.

RLEACH [19] is a secure routing protocol for cluster-based WSNs, using group key management, was proposed by Zhang et al. to solve the problem of secure LEACH. In this protocol, clusters are formed dynamically and periodically, can be thought as security extension of LEACH. RLEACH uses improve random pair-wise key management scheme (RPK), which use the one-way hash chain, symmetric and asymmetric cryptography to ensure security in LEACH. RLEACH resists to different attacks such as selective forwarding, sinkhole attacks, sybil attacks, and hello flood attack. Another advantage of RLEACH is that it is energy-efficient. In case, when a node transmit data to its CH, the member nodes among a cluster can close their wireless devices during the schedule creation phase or can sleep during the data transmission phase to save energy. Therefore, RLEACH balance the network security and the energy consumption in cluster-based WSNs.

EECBKM [20] Energy-Efficient Cluster Based Key Management is a cluster based technique for key management in WSNs. In this protocol, initially the clusters are formed in the network and the CHs are selected based on the energy cost, coverage and processing capacity. An EBS key set is assigned by the sink to every CH and cluster key to every cluster. The EBS key set contains the pair-wise keys for intra-cluster and inter-cluster communication. The data is made to pass through two phases of encryption during data transmission towards the sink. In this way security is ensured in the network. These keys are distributed to the nodes by the CH prior to communication. Secure channel is established between the nodes and the CH after the key distribution. Results have shown that this proposed technique reduces node-capture attacks and efficiently increases packet delivery ratio with reduced energy consumption.

SHEER [21] is a secure hierarchical energy-efficient routing protocol proposed by Ibriq and Mahgoub, which provides energy-efficient and secure communication on the network layer. For key distribution and authentication, securing the routing mechanism, SHEER uses HIKES (Hierarchical Key Establishment System) and also implements a probabilistic transmission mechanism to improve the network energy performance and lifetime.

SHEER defends the network against hello flood attack, sybil attack. The sinkhole attack will also fail because the attacker does not possess all keys, required for authentication. SHEER fail to protect the network from selective forwarding attacks.

SecLEACH [22] is a protocol for securing node-to-node communication in LEACH-based networks. Using random key pre-distribution, SecLEACH provides security in LEACH, introduced symmetric key and one-way hash chain to provide different performance numbers on efficiency and security depending on its various parameter values.

SecLEACH is an improvement of SLEACH [23], the first study in homogeneous WSNs focused on adding security to cluster-based communication protocols with resource constrained sensor nodes. SecLEACH provides authenticity, confidentiality, integrity and freshness for node-to-node communication. Otherwise SecLEACH is vulnerable to key collision attacks and do not provide full connectivity. The overheads in SecLEACH were computed using the number of CH value in the network which decrease the total energy consumption, and prolong the network's lifetime.

SS-LEACH [24] is a protocol based on LEACH protocol, considering routing security and network lifetime. Improving the method for electing CHs, the SS-LEACH protocol forms dynamic stochastic multipath CHs chains using nodes self-location technology and key pre-distribution strategy. So the SS-LEACH protocol strongly improves the energy-efficiency and hence prolongs the lifetime of the network. The SS-LEACH protocol can prevent compromised node and preserve the secrecy of the packet. It also can avoid sybil attack, selective forwarding and hello flooding.

NSKM [25], a Novel Secure Key Management module for Hierarchical Clustering WSNs provides an efficient scalable post-distribution key establishment that allows the hierarchical clustering topology platform to provide acceptable security services. In NSKM, there are three categories of keys; pre-deployed keys, network generated keys and the BS broadcasted keys. This module is the first implemented security module for WSNs that provides reasonable resistance against replay and node capture attacks. This work couples hierarchical clustering based routing with NSKM module. The selection of SCH among CHs is based on its location and its distance to BS. Most of communication types in WSNs have unique features of this work, using in-network keys generation and blending. The NSKM module is energy-efficient, has strong flexibility against susceptible attacks on WSNs, keeping the resource starved nature of sensor nodes. NSKM also ensures that the whole network is never compromised even if there has been an attack in the network. Furthermore, it is highly lightweight and scalable and is acceptable to be used in large WSNs.

AKM [26] is an Authenticated Key Management scheme for hierarchical networks based on the random key pre-distribution. Security is provided by using two kinds of keys, a pair-wise shared key between nodes, and a network key. To divide nodes into clusters, AKM scheme use an existing ring structure energy-efficient clustering architecture (RECA). Using more than one encryption key, AKM provides multiple level of encryption, secure cluster formation algorithm and avoid node captures. AKM provides confidentiality, global and continuous authentication of nodes in the network by periodically refreshing the network key. In general AKM scheme can be applied for different energy-efficient data dissemination techniques for sensors networks. However, if adversary re-enters the compromised node into the network before refreshing the current network key, the resiliency of AMK scheme will be same as given in Eschenauer et al. [11].

SRPSN [27] is a Secure Routing Protocol for Sensor Networks consists of a hierarchical network with CHs and cluster member nodes. CHs route the messages from sensor nodes. A preloaded symmetric key is shared between all CHs and the BS to protect data. SRPSN is also designed to safeguard the data packet transmission on the sensor networks under different types of attacks. A group key management scheme is proposed, which contains group communication policies, group membership requirements and an algorithm for generating a distributed group key for secure communication. Every sensor node contributes its partial key for computing the group key. One drawback associated with this protocol is that there is no authentication in the mechanism. Hence, SRPSN fail to protect against attacks like spoofing, altering, replaying. If the adversary uses the sybil attack, the problem will be more severe. The malicious node can also become a sinkhole. Another problem of this scheme is that children nodes will select a largest NBR node to relay data. However, energy consumption will be increased in this case.

SecRout [28], a Secure Routing Protocol for sensor networks is proposed by Yin and Madria to provide security against attack from compromised nodes in sensor networks. SecRout can detect if packets are dropped or modified by malicious nodes. In the SecRout protocol, only high efficient symmetric cryptography is used to secure messages, and the partial routing path is recorded in sensor nodes memory. Further, SecRout uses two types of keys: the master shared key used between the sink and CHs, and the cluster key among the clusters to encrypt the message. In SecRout all messages will be verified through MAC. It ensures that the messages received are not tampered, hence guarantees freshness. In SecRout, two-level architecture can greatly lower the message overhead. Therefore, SecRout can greatly save the energy, and decrease the usage of memory and bandwidth.

IKDM [29] is an Improved Key Distribution Mechanism, based on hierarchical network architecture and bivariate polynomial-key pre-distribution mechanism. In IKDM, each sensor has a unique id in the network. An offline Key Distribution Server (KDS) first initializes sensors before deployment by giving each sensor node a polynomial share. In order to setup a pair-wise key between two sensor nodes, they exchange their node ids first, and then nodes evaluate their stored polynomial. Since, sensors nodes can obtain the same value from the two distinct calculations, which can be used as their pair-wise communication key. Note that in IKDM, two communicating parties can establish a unique pair-wise key between them. IKDM scheme can achieve better network resilience against node capture attack, hence can provide efficient security and is not affected by the number of compromised sensors. IKDM scheme provides better scalability, network throughput, fixed key storage overhead, full network connectivity and is suitable for large-scale WSNs. Therefore IKDM scheme is more energy-efficient due to the lower communication overhead for sensor nodes during the pair-wise key establishment process.

## 8. Security Analysis

We describe some secure hierarchical routing protocols selected based on security mechanisms, security requirements, various routing attacks and performance metrics.

Security requirements for several routing protocols are summarized in Table 5. We observe that SecLEACH, SHEER, EECBKM, AKM, IKDM and NSKM address all the listed security requirements (authenticity, confidentiality, freshness and integrity) thus they are more secure than rest of the protocols if the security requirements is taken as criteria. According to the security requirements, selected protocols classification show that authentication and integrity are the most satisfied.

An overview of routing attacks in WSNs is shown in Table 6. From the table, it is clear that certain schemes defeats or mitigate the effect of various routing attacks. Considering the resistance against the routing attacks, Table 6 shows

Table 5: Security requirements for secure hierarchical routing protocols

| Secure Hierarchical Routing Protocol | Security Requirements | | | |
|---|---|---|---|---|
| | Authenticity | Confidentiality | Integrity | Freshness |
| RLEACH | Yes | | Yes | |
| EECBKM | Yes | Yes | Yes | Yes |
| SHEER | Yes | Yes | Yes | Yes |
| SLEACH | Yes | | Yes | |
| SecLEACH | Yes | Yes | Yes | Yes |
| SS-LEACH | Yes | Yes | | |
| NSKM | Yes | Yes | Yes | Yes |
| AKM | Yes | Yes | Yes | Yes |
| SRPSN | Yes | Yes | Yes | |
| SecRout | Yes | | Yes | |
| IKDM | Yes | Yes | Yes | Yes |

'Yes' means that protocol can achieve that security requirement.

Table 6: Resistance of routing attacks for secure hierarchical routing protocols

| Secure Hierarchical Routing Protocol | Routing Attacks in WSNs | | | | | |
|---|---|---|---|---|---|---|
| | Selective Forwarding | Sink-hole | Worm-hole | Sybil | Hello Flood | Node Capture |
| RLEACH | Yes | M | | Yes | Yes | |
| EECBKM | | M | | Yes | Yes | Yes |
| SHEER | Yes | M | | Yes | Yes | |
| SLEACH | Yes | M | | | Yes | |
| SecLEACH | Yes | | | Yes | Yes | |
| SS-LEACH | Yes | | | Yes | Yes | M |
| NSKM | Yes | M | M | Yes | Yes | Yes |
| AKM | Yes | Yes | M | Yes | Yes | Yes |
| SRPSN | Yes | | | | | |
| SecRout | Yes | M | | M | | Yes |
| IKDM | Yes | M | M | Yes | Yes | Yes |

'Yes' means that protocol defeats the attack and 'M' means that protocol mitigates the effect of attack based on our pre-evaluation.

that RLEACH, NSKM, EECBKM, AKM, SecRout, and IKDM are more resistant to routing attacks than rest of the secure protocols.

The detailed comparison results are summarized in Table 7. We observe that energy efficiency depends strictly on the communication overhead. Therefore, all schemes with lower communication overhead achieve energy efficiency (e.g. IKDM, SHEER, SS-LEACH, etc.). This is due to the fact that communications consume much more energies than the code execution or computation in WSNs. We also remark that approaches based on probabilistic key distribution (e.g. SLEACH, Sec-LEACH, etc.) are less energy-efficient than other schemes (e.g. IKDM based on deterministic approaches). Due to the fact that approaches based on probabilistic key distribution generate a lot of messages, require much more memory space. In contrast, the deterministic key distribution requires more computation time for nodes. Note that in WSNs, computation consumes less energy compared to the exchange of messages between sensor nodes. Considering the scalability, Table 7 also shows that probabilistic approaches are less scalable than other schemes.

Table 7: Comparison summary based on security mechanisms, performance and efficiency of some selected secure hierarchical routing protocols implemented for WSNs.

| Protocol Name | *A Comparative Overview Of Representative Secure Hierarchical Routing Protocols For WSNs* | | | | | | | | | |
|---|---|---|---|---|---|---|---|---|---|---|
| | *Ref* | *Cryptography Scheme* | *Key distribution and Management Scheme* | *Authentication Scheme* | *Storage Load* | *Comm. Load* | *Scalability* | *Robustness* | *Connectivity* | *Energy Efficiency* |
| RLEACH | [19] | Symmetric key cryptography | Improved Random pair-wise key management (IRPK) | Authentication is achieved via IRPK | High | Medium | Good | Good | Probabilistic | Medium |
| EECBKM | [20] | | EBS-based key Management schemes | Via Key Management | Low | Low | Medium | Good | 100% | Good |
| SHEER | [21] | Symmetric key cryptography | Hierarchical key management and authentication scheme | Authentication is achieved via HIKES | Medium | Low | Good | Good | 100% | Good |
| SLEACH | [23] | Symmetric key cryptography | | MAC | High | Medium | Medium | Limited | Probabilistic | Medium |
| Sec-LEACH | [22] | Symmetric key cryptography | Random key pre-distribution scheme | Don't provide broadcasts authentication | High | Medium | Medium | Limited | Probabilistic | Medium |
| SS-LEACH | [24] | Symmetric key cryptography | Keys pre-distribution strategy | | Medium | Low | Medium | Limited | 100% | Good |
| NSKM | [25] | | Key management schemes based | MAC | Low | Low | Good | Good | 100% | Good |
| AKM | [26] | | Random Pre-distribution Key Management | Via Key Management and MAC | High | Medium | Good | Good | Probabilistic | Medium |
| SRPSN | [27] | Symmetric key cryptography | Group key management scheme | MAC | Medium | Low | Medium | Low | 100% | Good |
| SecRout | [28] | Symmetric key cryptography | The Scheme introduced in LEAP [15] | MAC | Low | Low | Good | Limited | 100% | Good |
| IKDM | [29] | | Bivariate polynomial-key pre-distribution mechanism | Via polynomial key pre-distribution mechanism | Low | Low | High | Good | 100% | Good |

There exist some surveys on secure hierarchical WSNs [30]. However none of them address the energy consumption constraints following the security mechanisms.

Based on hierarchical topology, Sharma and Jena [30] consider that all selected works are energy efficiency. However, they do not pay much attention on energy constraints when different security mechanisms are used. This is very crucial because technique based probabilistic and deterministic don't have the same impact on energy consumption. In addition, Sharma and Jena [30] did not address the performance requirements study (e.g. memory overhead, computation overhead etc.), which is more important because it is strictly bound to the energy consumed.

In the paper, we presented an overview of well-known routing protocols for WSNs and a technical overview of each protocol. We also provide a comprehensive and informative comparison of them, which we believe is a significant improvement, when compared to other comparative studies in WSNs.

## 4. Conclusions and Future Research

The main goal of a routing protocol design is to provide energy efficiency and extend network lifetime. Sensor nodes are susceptible to a number of routing attacks depending on the nature of the WSNs, the limited memory resources and energy constraints. In order to provide security in WSNs and mitigate the security threats to routing protocols, secure routing protocols to be used. In this paper, we reviewed and analysed some secure cluster-based routing protocols. The comparative study show that some selected schemes can well balance between security level and the associated energy consumption overhead. An informative overview of protocols is given and their advantages and disadvantages listed. We also presented detailed comparison based upon various criteria in the analysis section. Further, research would be needed to address issues related to secure routing under the mobility for resource constrained WSN. The study may help to orient the development of future proposals well adapted in the area of security issues in routing protocols for WSNs.

**Acknowledgments**

This work was supported by the High-tech Research and Development Program of China (Grant No.2011AA040101-3).